\documentclass[]{cJAS2e}

\usepackage[pdftex]{hyperref}
\usepackage{color}

\begin{document}
\doi{10.1080/02664760802715922}
\issn{1360-0532}
\issnp{0233-1934}
\jvol{36} \jnum{10} \jyear{2009} \jmonth{October}

\markboth{G. K. Skinner and G. H. Freeman}{Journal of Applied Statistics}


\title{Are soccer matches badly designed experiments? }

\author{G. K. Skinner$^{\rm a}$$^{\ast}$\thanks{$^\ast$Corresponding author. Email: skinner@milkyway.gsfc.nasa.gov}
 and G. H. Freeman$^{\rm b}$\\\vspace{6pt}  
 $^{\rm a}${\em{Dept. of Astronomy, University of Maryland, College Park, MD 20742, USA  \& CRESST \& NASA-GSFC, Greenbelt Road, Greenbelt, MD 20771, USA}};\\
  $^{\rm b}${\em{University of Warwick, Coventry, UK}}\\\vspace{6pt}
  \received{Submitted 10 March 2008} 
  }

\maketitle

\begin{abstract}
Models in which the number of goals scored by a team in a soccer match follow a Poisson distribution, or a closely related one, have
been widely discussed. We here consider a soccer match as an experiment to assess which of two teams is superior and examine the
probability that the outcome of the experiment (match) truly represents the relative abilities of the two teams.
 Given a final score it is possible by using a Bayesian approach to quantify the probability that it was or was not the
 case that `the best team won'. For typical scores, the probability of a misleading result is significant.
  Modifying the rules of the game to increase the typical number of goals scored would improve the situation, but a level of confidence that
  would normally be regarded as satisfactory could not be obtained unless the character of the game were radically changed.\\
\vspace{5mm}
\\
Note: For copyright reasons this is a preliminary form of an article whose final and 
definitive form has been published in {\it Journal of Applied Statistics} {\bf 36,} 1087--1095 (Oct 2009) 
[copyright Taylor \& Francis] and is
available online at: \href{http://www.informaworld.com/smpp/content~db=all~content=a915275365}{\textcolor{blue}{http://www.informaworld.com/smpp/content/$\sim$db=all$\sim$content=a915275365}}.
Contact the corresponding author in case of difficulty in accessing the published paper.

\bigskip

\end{abstract}

\section{Introduction}

The distribution of the number of goals scored in association football (soccer) matches has been investigated
by various authors over the last half century \citep{moroney56,maher82,dixon98,greenhough01,bittner07}.
The emphasis has usually been on finding models that describe the distributions observed in large data-bases of
match results, often with the objective of forecasting results, of optimising
playing or betting strategies, or of studying the efficiency of the betting market. We here use the results of this work
not for forecasting but to consider what, if anything, can be deduced from the result of a match
about the relative strengths of the teams.

A football match can be regarded as an experiment to determine which of the two teams is in some sense
superior, or  perhaps one should say ``.. is superior given
the date and circumstances of the match". The statistical models of goal numbers that have been developed
have major implications for the probability that the experiment gives a correct result -- that is to say that
``the best team won". These have not been widely discussed, and here we quantify them and extend some of the considerations
to tournaments involving many teams and matches.

In the simplest models which have been considered, goal scoring is regarded as a Bernoulli process in which
the probability of team $A$ scoring in time interval $dt$ is $\lambda_a dt$, where $\lambda_a$ is constant,
and similarly that for team $B$ is $\lambda_b dt$. This leads to the probability of the result $(N_a,N_b)$
being given by the product of two univariate Poisson distributions :
\begin{equation}
 \wp\left\{ (N_a,N_b)|(\alpha_a,\alpha_b) \right\} \; = \; {{\alpha_a^{N_a} \exp(-\alpha_a)}\over{N_a!}}\:{{\alpha_b^{N_b} \exp(-\alpha_b)}\over{N_b!}}
 \; = \;{\alpha_a}^{N_a} {\alpha_b}^{N_b} {{\exp(-\alpha_a-\alpha_b)}\over{N_a!N_b!}}
 \label{conditional_prob_2_univariate}
\end {equation}
with expectation values $\alpha_a = \lambda_a T$ and $\alpha_b = \lambda_b T$, where $T$ is the match duration.



In practice the $\lambda$ are not constant. Variation of $\lambda$ during a match would not by itself
invalidate the Poisson model as a mean level can be used. The well-known ``home team advantage" implies that
 $\lambda$ is likely to depend on where the match is played \citep{lee77,clarke_norman95}.
 This is sometimes accommodated by analysing separately
 the score at home and away matches. More difficult to handle is the fact that it might be expected that for psychological
or strategic reasons $\lambda$ might depend on the number of goals already scored by either or both of the two teams.
There is statistical evidence that this is indeed the case.

In considering the distribution of goal scores at an aggregate level, it was noted from an early stage that
there is an excess of high scores compared with a Poisson distribution. Maher \cite{maher82}
pointed out that the negative binomial distribution used by Reep et al. \cite{reep71} and, implicitly,
by Moroney \cite{moroney56} to provide a better description of the tail of the distribution can be regarded as the weighted sum of
Poisson distributions with different means. Thus it is consistent with the expected effect of including
 results obtained with different $\alpha$ in the aggregate.

Greenhough et al. \cite{greenhough01} found that the high score tails in some datasets could not be modelled by either Poisson or
negative binomial distributions and were better described by using extremal statistics. Bittner et al.  \cite{bittner07}
explain the excess in terms of a dependance of $\alpha$ on the number of goals already scored -- a dependence that
they ascribe to `football fever' -- a self affirmation in which goals encourage more goals. This effect appears to dominate over one in which a winning team either relaxes or plays a purely defensive game.

It is obviously a simplification to model each teams score independent of the other. Modifying the simple
univariate Poisson model of equation (1) to allow for a correlation between the two scores leads
to a bivariate Poisson distribution for $\wp(N_a,N_b)$. Maher \cite{maher82} used a bivariate Poisson model to correct
the tendency of simpler models to underestimate the number of draws. Lee \cite{lee99} has discussed
such models in the context of Australian rugby league scores and compared them with others.
Crowder et al \cite{crowder02} have applied them to football results, and  Karlis and Ntzoufras \cite{karlis03} to
both football and water polo. In some of their models  Bittner et al. 
\cite{bittner07} 
allow for the correlation  by making the
scoring rate depend on the number of goals scored by \textbf{both} teams, potentially in different ways.
Karlis  and Ntzoufras \cite{karlis05} developed an inflated bivariate Poisson distribution to take account simultaneously of both correlations and non-Poissonian tails.

We here consider the level of confidence that, given the statistical uncertainty implied by models such as those discussed above,
 one can have in the outcome of a match.  First (in Section 2) the simple model of Equation 1 is used. In Section 2.2 we show that
 the conclusions are little changed by the use of more sophisticated models. Section 3 examines the implications for a tournament
 involving a series of matches.

\section{Level of confidence in the outcome of a match}

If a football match is a well designed experiment the winning team -- that which has scored the greatest number of goals at
the end of the match -- will be the one with a higher level of skill. By making certain simplifying assumptions the
probability that the experiment gives the wrong result for purely statistical reasons can be quantified.
When considering the outcome of a single match, many of the issues which complicate the analysis of aggregate
scores can be ignored. We will put to one side issues of whether a team has lost its form, changed its manager
or is at home or away, and we will consider that the experiment has led to the correct result if the team
that is stronger, on the day and in the particular circumstances of the match, wins.
If it were possible to replay a match many times in exactly the same circumstances then after a sufficient number of matches one team could eventually be shown to be
superior to the other, with whatever level of confidence was required, but for some fraction of individual matches the score would imply a reversed ranking.

We do not in practice know $\alpha_a, \alpha_b$ but after the match we know the final score, ($N_a, N_b$). Given the number of goals scored by each team
and assuming that each follows a Poisson distribution independent of the other, then equation \ref{conditional_prob_2_univariate} allows the probability
$ \wp \left\{ (N_a, N_b) | (\alpha_a, \alpha_b)\right\}$ to be found as a function of $\alpha_a, \alpha_b$, but we are more
concerned with $ \wp \left\{ (\alpha_a, \alpha_b) | (N_a, N_b)\right\} $.

Bayes' theorem allows us to write
\begin{equation}
  \wp \left\{ (\alpha_a, \alpha_b) | (N_a, N_b)\right\}
  = \wp\left\{ (N_a, N_b) | (\alpha_a, \alpha_b)\right\} {{\wp\left\{ \alpha_a, \alpha_b\right\} }\over {\wp\left\{ N_a, N_b\right\}}}
\end{equation}

We will initially assume no prior knowledge about the strength of the teams. This means that, before the match, any combination of $\alpha_a, \alpha_b$ is equally
likely, or in other words that the prior probability ${\wp\{\alpha_a, \alpha_b\}}$ is constant. For a given result $N_a,N_b$,
Equation \ref{conditional_prob_2_univariate} 
then also gives the relative probability $\wp\left\{ (\alpha_a, \alpha_b) | (N_a, N_b)\right\}$.

In fact the objective of the experiment is only to know which team is superior, that is to say whether $\alpha_a > \alpha_b$ or $\alpha_a < \alpha_b$.
The convention in football and most games is the Bayesian one -- one adopts the solution that has the highest probability
of producing the observed result. In the absence of prior information, the case that is most likely to lead to the result $(N_a,N_b)$ is
 $\alpha_a = N_a, \alpha_b = N_b$, so if $N_a > N_b$ then we deduce that  $\alpha_a > \alpha_b$ and we declare team $A$ to be superior.
But a range of solutions surrounding the best one is also allowed.
To find the probability $w$ that the result does not correctly reflect the abilities of the teams, we need to integrate
over the relevant part of $\alpha_a, \alpha_b$ space. For $N_a<N_b$
\begin{equation}
   w(N_a,N_b) \; = \;\wp\left\{ (\alpha_a > \alpha_b) | (N_a, N_b)\right\}  =  { \int_0^\infty \int_{\alpha_b}^\infty  \wp\left\{ (N_a, N_b) | (\alpha_a, \alpha_b)\right\} d\alpha_a d\alpha_b  }
\end{equation}

\subsection{If two unknown univariate Poisson teams play each other...}

\begin{figure}[h]\centerline{\rotatebox{-90}{\scalebox{0.5}{\includegraphics{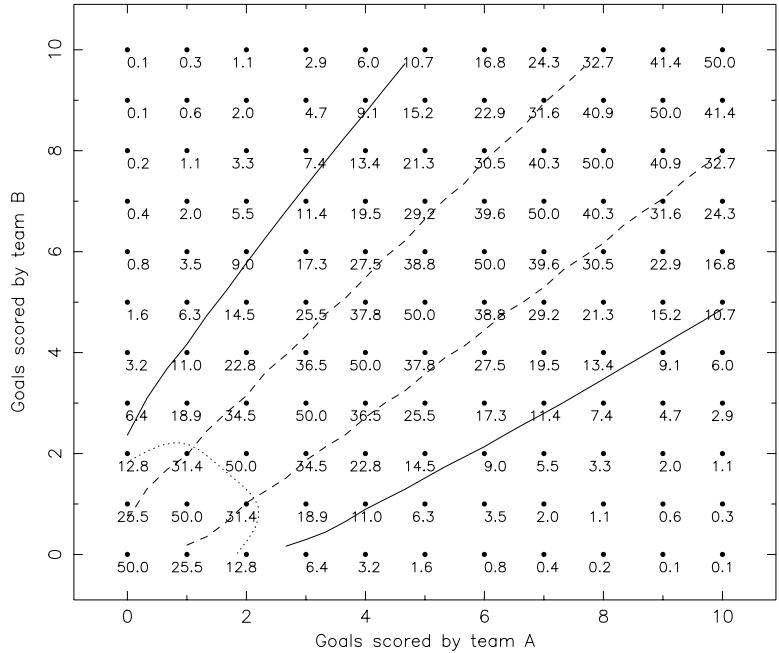}}}}
 \caption{ The probability $W$ that a particular result $N_a,N_b$ does not correctly represent the relative abilities of the two teams. Calculated with
 flat prior probability functions. The probabilities are meaningful only for integer numbers of goals, but interpolated contours are shown to define
 zones in the plot corresponding to $W<10$\% (continuous lines) and to $W<32$\% (corresponding approximately to a 1$\sigma$ result). The dotted line encloses 50\%
 of the results in FIFA world cup matches}
 \label{fig_flat}
 \vspace{0.1cm}
  \end{figure}

Suppose we have any model that gives a probability of different scores as a function of a pair of expectation values $\alpha_a, \alpha_b$
 (or of some other parameters characterising the two teams). For a given final score, we can now evaluate the
 probability that the match (experiment) gave a correct or a misleading result.
Figure \ref{fig_flat} gives results using the simple univariate Poisson model of Equation \ref{conditional_prob_2_univariate}. It can be see that
the probability of a false result is considerable unless the goal difference is very high. For differences less than 3--4 goals the result lacks
the 90\% confidence which within quantitative disciplines is frequently considered a minimum acceptable level of confidence in the outcome of an experiment.
The majority of final scores that occur in top quality football fail to reach even `1-sigma' confidence.

\subsection{More complex models}

\begin{figure}[h]\centerline{\scalebox{0.6}{\includegraphics{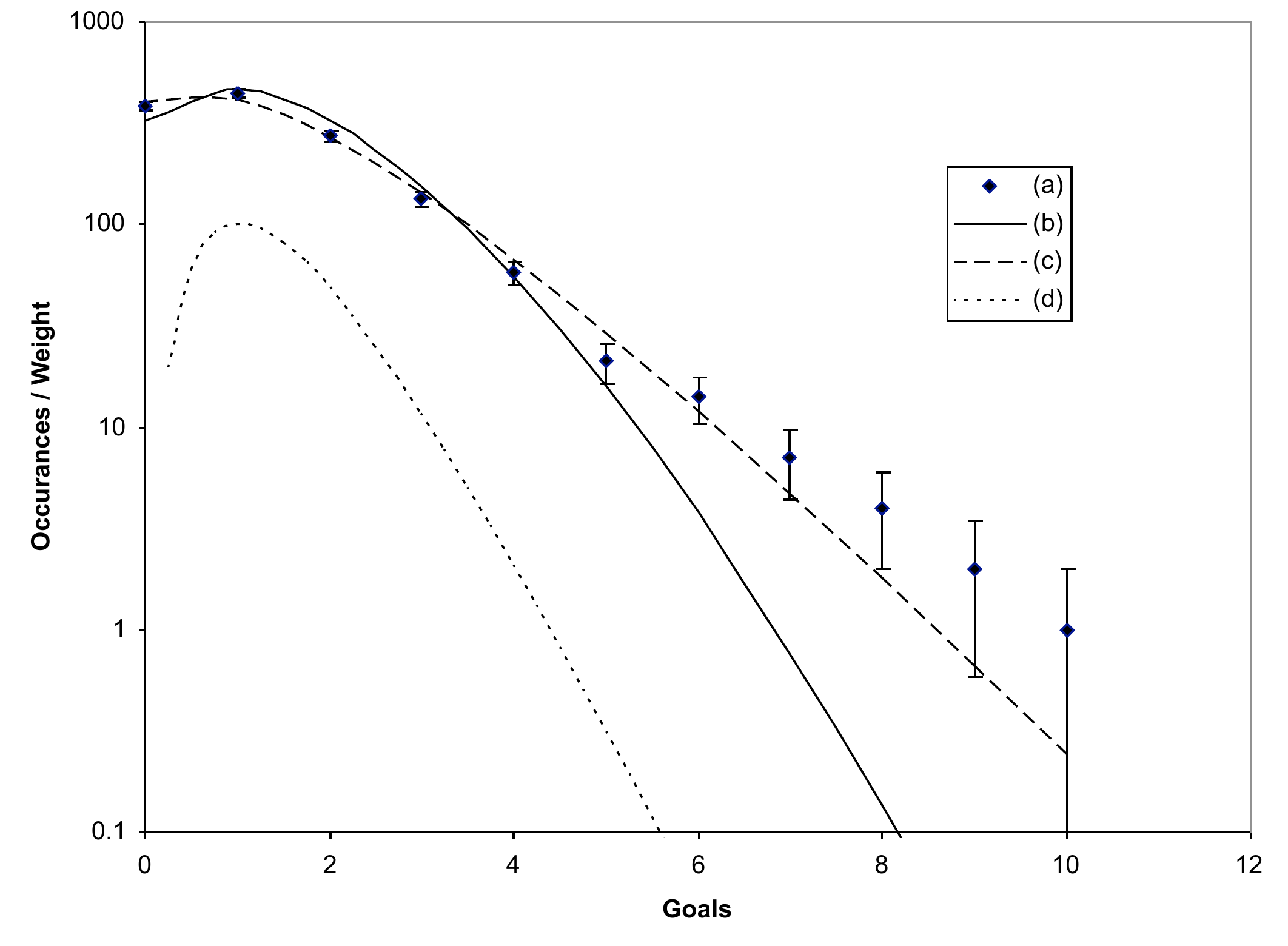}}}
 \caption{ (a) The distribution of the number of goals per match scored by teams in the FIFA world cup 1938-2006. (b) A Poisson distribution with the same mean (1.43).
 (c) A negative binomial fit. (d) The distribution of the expectation values of Poisson distributions which would have to be combined to produce (c)
 (normalised to a maximum of 100).}
 \label{fits}
 \vspace{0.1cm}
  \end{figure}

As has already been discussed, the use of univariate Poisson distributions for the two teams is an approximation.
We use here as an example the final scores during the FIFA world cup series 1938-2006 (after extra time where applicable, but without penalty shoot-outs).
 The distribution of number of goals scored is shown in Figure \ref{fits}. There is an  excess  of high scores compared with a
 Poisson distribution having the same mean (b), as seen by many authors in other datasets. A better fit is provided by a negative binomial distribution (c)
 with parameters adjusted to maximise the likelihood, though there are still indications of a slight excess of high scores.
 Re-evaluating the data shown in Figure \ref{fig_flat} with the negative binomial fit changes the values very little.

Strictly, the inclusion of results after extra time must have some effect.  For example,
the scores will not be Poissonian if a decision to prolong the match depends on the score.
Thus some small part of the ``supra-Poissonian" variance must be due to including data from extended duration matches.
The effect of extra time in those matches where extra time was played is to reduce the fraction of drawn matches
from about 25\% to 12.3\%. However, the impact on the data in Figure \ref{fits}
of using results after normal time rather than after extra time is to shift the points only
by less than the size of the symbols.

More importantly, the assumption of a uniform prior is obviously invalid -- we know that there are no teams around
that regularly score thousands of goals per match!
The distribution of $\alpha$ must actually be rather narrow, otherwise analyses of large databases would not
find even approximately a Poisson score distribution. If we use narrower prior probability distributions for
$\wp\left\{\alpha_a\right\}$, $\wp\left\{\alpha_b   \right\}$ (keeping them the same for $A$ and $B$, because we want to start
the experiment with no presumption about the outcome)  the significance
which should be attached to the outcome of a match will be further reduced.  The experiment is trying to differentiate between two teams
already known to be close in ability.

\begin{figure}[h]\centerline{\rotatebox{-90}{\scalebox{0.5}{\includegraphics{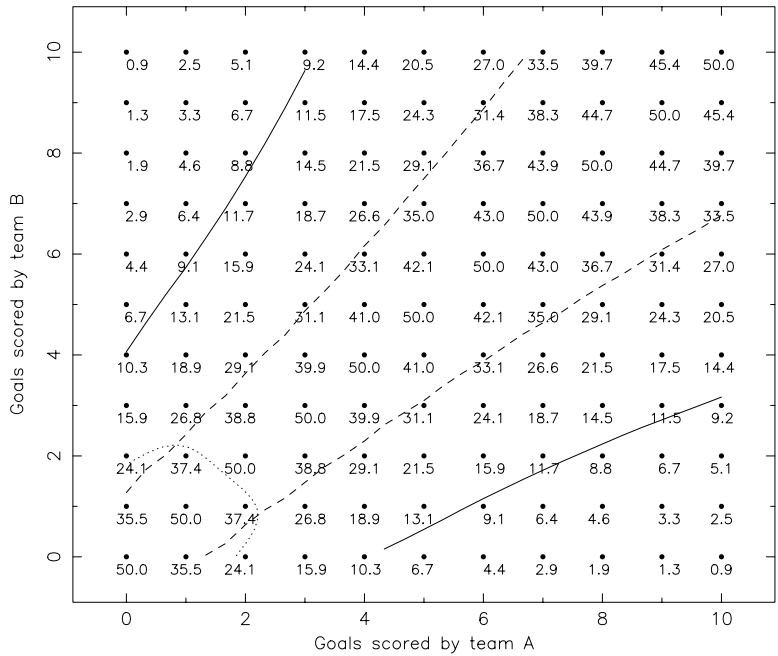}}}}
  \caption{ As figure \ref{fig_flat} but with a prior probabilities following the form of curve (d) in Figure \ref{fits}}
 \label{fig_prior}
 \vspace{0.1cm}
  \end{figure}

The negative binomial distribution can be expressed as a weighted mixture of Poissonian ones :
\begin{equation}
  f(n) = {\Gamma(r+n) \over{n! \Gamma(r)}} p^r (1-p)^k = \int_0^\infty  Poisson(n|\alpha)Gamma\left(\alpha| r,(1-p)/p\right) d\alpha
\end{equation}
Figure \ref{fits}(d) shows the Gamma distribution describing the decomposition of (c) into Poissonians with different expectation values.
This can be interpreted as showing the intrinsic range of $\alpha$ values. Using a prior of this form increases the probability $w$ of a misleading result,
 as seen by comparing Figure \ref{fig_prior} with Figure \ref{fig_flat}.
If some of the high score tail is due to `goal fever' or other effects such as the general downward drift in mean scores over the 58 years covered
by the data, then the spread in $\alpha$ will be even narrower. Thus two teams playing each other are likely to be even closer in ability  and the match outcome even more uncertain.

We have considered the possibility of using bivariate distributions, but for this dataset there appears to be no
correlation between the scores of the two teams and so there is no reason to do so.

\section{The situation for a tournament}

Tournaments are sometimes organised such that the fate of a team does not depend on a single match but that they have the
possibility of compensating a bad  result by other good ones. In this way, by performing multiple experiments,  the statistical significance of the
outcome can be increased. On the other hand it is frequently the case that the eventual winner has to pass through
many eliminating rounds, increasing the probability of error. Some studies of tournament design have considered the effects of the unreliability
of the result of a single match and how to maximise the probability that the best team/player goes forward to the next round or wins.
Most work of this sort has assumed that Gaussian noise is introduced into a comparison  process, often in the context of tournaments (such as Wimbledon)
where a ranking or seeding of the competitors is used in selecting pairings ({\it e.g.} \cite{adler_etal95,glickman08}).
In soccer, as we have seen, the statistics are close to Poissonian and in the FIFA World Cup used an example here,
in recent series for the first round of the final competition teams have been grouped into ``little leagues" of 4 teams
using some degree of seeding but in combination with a random draw.

Multi-match tournaments offer an opportunity for verifying some of the ideas discussed here.
Often all combinations of group of teams play each other. If the result of each match provided a valid comparison
of the relative abilities of the two teams, the situation that A beats B beats C beats A should never arise. We refer to this
as an intransitive triplet. Note that up to this point it has only been assumed that the relative ability of two teams at a
particular time and in particular circumstances is to be tested.
We now have to imagine that a teams ability does not change and that there is a real sense in which one team may be superior to another.
But even in the absence of changes, an anomalous combination of results can arise. If the true ranking is $A>B>C$ but the outcomes of each
of the 3 matches has probability  $w$ of not corresponding to that ranking, then there is a probability $w(1-w)$
that an intransitive triplet will result.

The FIFA world cup results provide a database which includes 355 examples of triplets.
Of the 147 which do not involve a drawn match, 17 (12\%)
are intransitive. This seems comparatively low, but we note that even  if the match outcomes were entirely random the fraction expected would only be 25\%.
An approximate estimate of the number which might be expected can be obtained by noting that the scorelines of the non-drawn
matches in this database have uncertainties averaging $w=$20.0\% (here and in the discussion which follows,
values from Figure \ref{fig_flat} have been used as they are the most optimistic).
This   corresponds to $w(1-w)=$ 16\% or 23.5$\pm$4.8  intransitive triplets expected, reasonably consistent with the 17 seen.

While multiple combinations of teams playing each other can reduce the uncertainty in the outcome, like many other
competitions the final stages of recent FIFA world cup series involve a knockout.  16 teams are reduced to 1 in 4 stages. Even if the best team reaches the
16, if it is to gain the cup it must avoid a false result in all 4 of its last games. As draws are resolved by a penalty shootout, which may be treated as
nearly random, the appropriate mean value of $w$ is that including draws, which is not 20\%, 
but 27\%\def\thefootnote{\dagger}\footnote{We note that in a tournament $w$ may not be constant, but may increase in later stages as teams become more equally matched. For simplicity we adopt a mean value.}.
The best team has only a probability of about 28\% of winning the cup, even it reaches the last 16. For the actual match scores which led Italy through its last
4 matches  to the 2006 cup the corresponding value comes to 30\%.

\vspace{5mm}

\section{Conclusions}

It is apparent from Figures \ref{fig_flat} and \ref{fig_prior} that the scores which most frequently arise correspond to relatively
high probabilities of a misleading outcome. In the recent FIFA World Cup only 5 matches among 64 had scores
corresponding to better than 90\% confidence in the result and one third had results which should be classified as `$<1\sigma$'.
Even on very optimistic assumptions there is less than one chance in three that it was the best team that won the cup.

 The possibility of increasing the size of football (soccer) goal mouths to make the game more interesting
 has been discussed and an attempt has been made to use somewhat dubious simple dynamics to quantify the likely
 effect of a specific change in goal size on the number of goals scored  \citep{mira06}. The present analysis cannot be
  used to estimate by how much the mean score would have to be increased to achieve a
  given level of confidence in the result without considering the likely difference in the level of
skills of the two teams. In principle one could imagine continuing the match with successive periods of extra time until the goal
difference becomes large enough to yield a chosen level of confidence. Such open ended matches would not
be popular with those planning television coverage (though the undefined duration of tennis matches is reluctantly accommodated).
In either case is clear that the character of the game would be entirely changed.

The sportswriter  Grantland Rice once wrote  \citep{rice}
{\sl 
\vspace{6mm}\\
``When the One Great Scorer comes to mark beside your name,\\
He marks -- not that you won nor lost --\\
but how you played the game."
 \vspace{6mm}\\
}
 Perhaps its just as well, for in soccer the one bears little relationship to the other.

\bibliography{Poisson_goals}
\bibliographystyle{cJAS} 

\end{document}